\documentclass[twocolumn,showpacs,preprintnumbers,amsmath,amssymb]{revtex4}

\usepackage{graphicx}
\usepackage{dcolumn}
\usepackage{bm}

\begin{document}
\title{Measurement of the Slope Parameter for the $\eta\rightarrow3\pi^{\circ}$ Decay 
in the $pp\to pp\eta$ Reaction}

\author{
M.~Bashkanov$^1$, D.~Bogoslawsky$^2$, H.~Cal\'en$^3$, F.~Capellaro$^4$,  H.~Clement$^1$, L.~Demir\"ors$^5$, 
C.~Ekstr\"om$^3$, K.~Fransson$^3$, L.~Gustafsson$^4$, B.~H\"oistad$^4$, G.~Ivanov$^2$,   M.~Jacewicz$^4$, 
E.~Jiganov$^2$, T.~Johansson$^4$,  S.~Keleta$^4$,  I.~Koch$^4$, S.~Kullander$^4$, A.~Kup\'s\'c$^3$, 
A.~Kuznetsov$^2$, P.~Marciniewski$^3$, R.~Meier$^1$, B.~Morosov$^2$, W.~Oelert$^6$, C.~Pauly$^5$, 
Y.~Petukhov$^2$, H.~Pettersson$^4$, A.~Povtorejko$^2$, R.J.M.Y.~Ruber$^3$, K.~Sch\"onning$^4$, W.~Scobel$^5$, 
T.~Skorodko$^1$, B.~Shwartz$^7$, V.~Sopov$^8$, J.~Stepaniak$^9$, V.~Tchernyshev$^8$\footnote{Deceased}, P.~Th\"orngren Engblom$^4$, 
V.~Tikhomirov$^2$, A.~Turowiecki$^{10}$, G.J.~Wagner$^1$, U.~Wiedner$^4$, M.~Wolke$^{4,6}$, A.~Yamamoto$^{11}$, 
J.~Zabierowski$^9$, J.~Z{\l}oma\'nczuk$^4$
}

\affiliation{
$^1${Physikalisches Institut der Universit\"at T\"ubingen \nolinebreak T\"ubingen,\nolinebreak Germany}\\ 
$^2${Joint Institute for Nuclear Research \nolinebreak Dubna,\nolinebreak Russia}\\ 
$^3${The Svedberg Laboratory, Uppsala, Sweden}\\ 
$^4${Institutionen f\"or K\"arn- och Partikelfysik \nolinebreak  Uppsala \nolinebreak University, \nolinebreak Uppsala, \nolinebreak Sweden}\\
$^5${Institut f\"ur Experimentalphysik~Universit\"at~Hamburg, \nolinebreak Hamburg,~Germany}\\ 
$^6${Institut f\"ur Kernphysik Forschungszentrum J\"ulich Germany}\\ 
$^7${Budker Institute of Nuclear Physics,\nolinebreak Novosibirsk,\nolinebreak Russia}\\
$^8${Institute of Theoretical and Experimental Physics,\nolinebreak Moscow,\nolinebreak Russia}\\  
$^9${Soltan Institute of Nuclear~Studies,~Warsaw~and~Lodz,~Poland}\\  
$^{10}${Institute of Experimental Physics,\nolinebreak Warsaw,\nolinebreak Poland}\\ 
$^{11}${High Energy Accelerator \nolinebreak Research \nolinebreak Organization,\nolinebreak Tsukuba,\nolinebreak Japan}\\
(CELSIUS-WASA Collaboration)  
}

\date{\today}
            
\begin{abstract}
The  CELSIUS/WASA setup  is used  to measure the  3$\pi^{\circ}$  decay of
$\eta$ mesons produced in $pp$ interactions with beam kinetic energies of
1.36 and  1.45 GeV. The  efficiency-corrected Dalitz plot  and density
distributions for  this decay  are shown, together  with a fit  of the
quadratic  slope  parameter $\alpha$  yielding  $\alpha  = -0.026  \pm
0.010(stat)  \pm  0.010(syst)$.   This  value is  compared  to  recent
experimental results and theoretical predictions.
 
\end{abstract}


\maketitle

\section{Introduction}

Isospin violating decays into three pions are two of the most probable
$\eta$ meson decay modes: $BR(\eta\to 3\pi^{\circ}) = 32.5\%$ and
$BR(\eta\to\pi^{\circ}\pi^+\pi^-)=22.6\%$ \cite{Yao:2006px}.  It
appears that contribution of electromagnetic processes is suppressed
\cite{Sutherland:1967vf,Baur:1995gc} and the decays are driven by an
isospin violating term of the QCD Lagrangian proportional to $m_d -
m_u$.  The partial width of the $\eta\to \pi^{\circ} \pi^+\pi^-$ decay
calculated using current algebra (PCAC) is 66 eV \cite{PhysRevLett.18.1170},
much below the experimental value 294$\pm$16 eV \cite{Yao:2006px}.
Second order contributions in the low energy expansion of the QCD Lagrangian (Chiral
Perturbation Theory -- CHPT) were calculated by Gasser and Leutwyler
increasing the result to 160 eV \cite{Gasser:1984pr}.  The big change
with respect to the first order calculations implies the importance of
$\pi\pi$ interaction in the final state.  An elegant method of
including the interaction up to higher orders is provided by
dispersion relations.  There are two calculations using this technique
\cite{Anisovich:1996tx, Kambor:1995yc} but employing different
formalism.  They lead consistently to an enhancement of the decay rate
by about 14\%.  A free parameter in this approach is the value of the so
called {\em subtraction point} and it was constrained using CHPT calculations.

The decay width can be expressed in the factorized form:
\begin{equation}
\Gamma=\left(\frac{Q_D}{Q}\right)^4\bar{\Gamma}
\end{equation}
where the dependence on the $m_d - m_u$ is contained only in the $Q$
term:
\begin{equation}
\frac{1}{Q^2} = \frac{m_d^2-m_u^2}{m_s^2 - \frac{1}{4}(m_d + m_u)^2}.
\end{equation}

Normally the calculations of $\Gamma$ are performed using the Dashen
theorem \cite{Dashen:1969eg} where $Q=Q_D\equiv 24.1$.  Since
$\Gamma$ is sensitive to the exact value of $Q$, it was suggested that
the decay might provide precise constrain for the light quark mass
ratios \cite{Leutwyler:1996qg}.  Namely $Q$ determines the major axis
of the ellipse in the $m_u/m_s$, $m_d/m_s$ plane.  One important
prerequisite is the reliability of the $\bar{\Gamma}$ calculations.  The
calculations can be tested by comparing the ratio $\Gamma(\eta \to
3\pi^{\circ})/\Gamma(\eta \to \pi^+\pi^-\pi^{\circ})$ and the density
distributions in the respective Dalitz plots with experiment.
Recently $Q$ was derived from preliminary KLOE data on
$\eta\to\pi^+\pi^-\pi^0$ decay \cite{Giovannella:2005rz,
Ambrosino:2007mq}, by determining the subtraction point within
the dispersion relation approach of \cite{Kambor:1995yc}, yielding
a $Q$ value 22.1 \cite{Martemyanov:2005bt}.

\begin{figure}
    \includegraphics[width=0.23\textwidth]{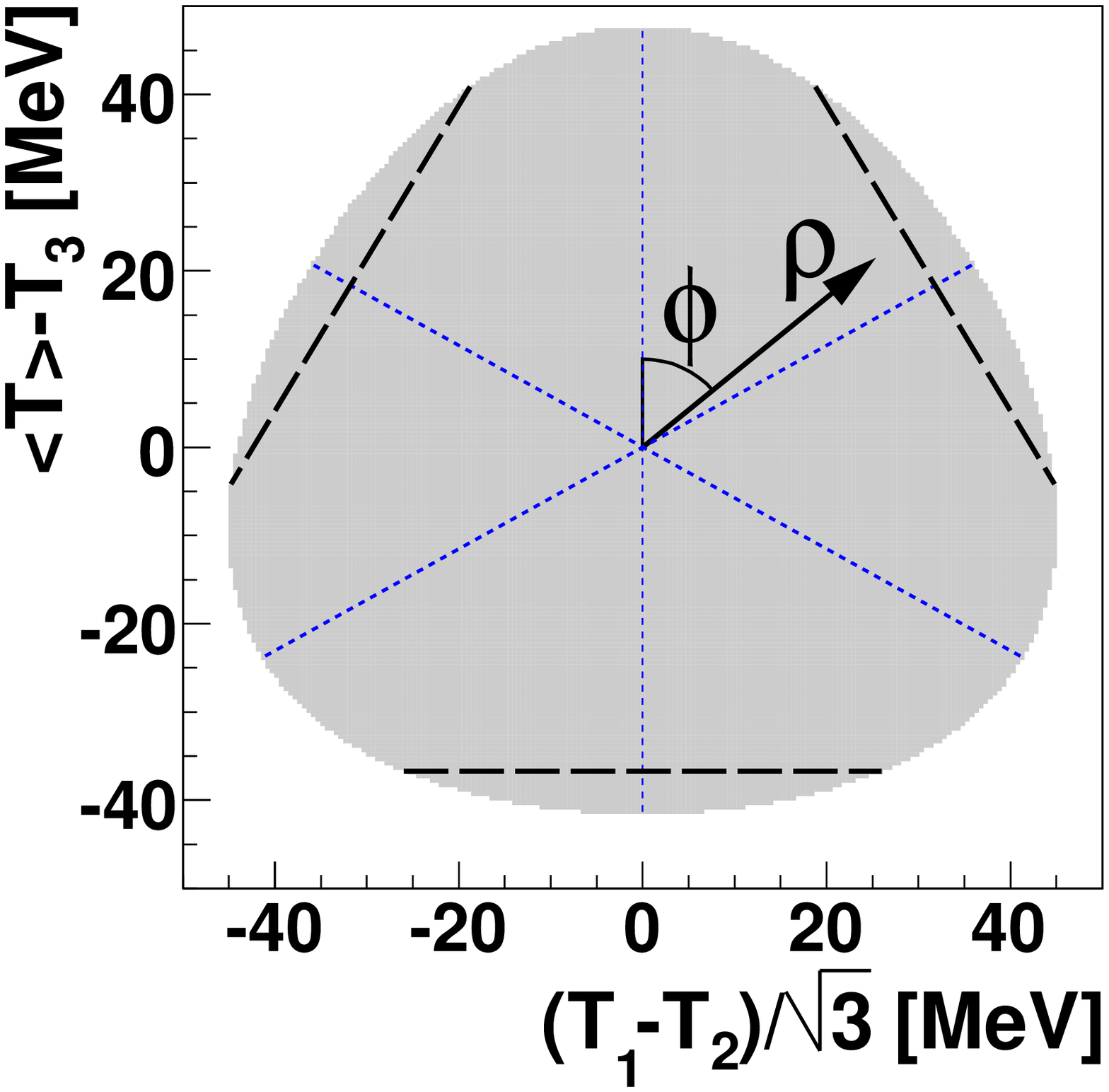}
    \includegraphics[width=0.24\textwidth]{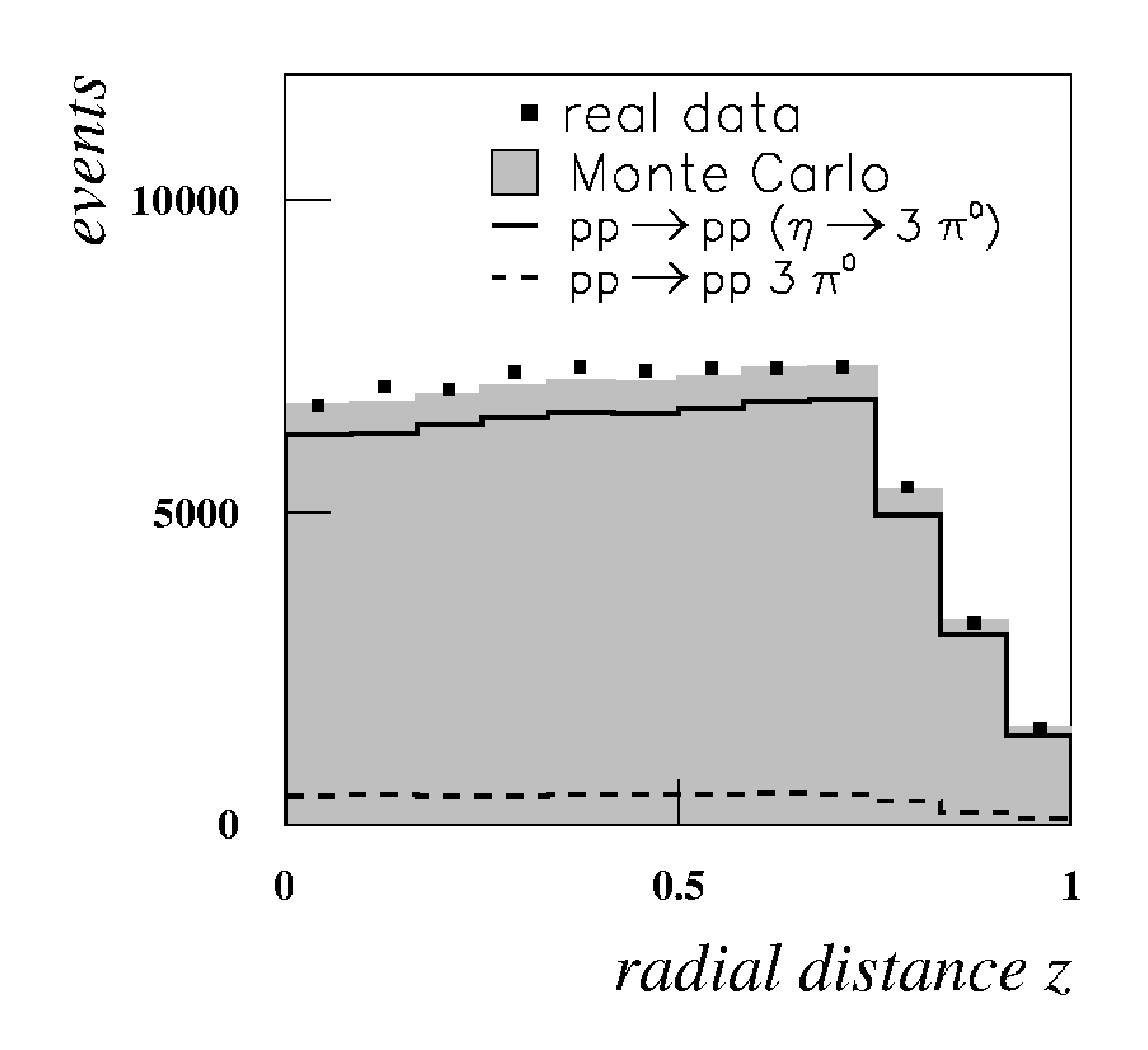}
  \caption{\label{fig:dalitzplot}
(Left) Symmetrized Dalitz plot for $\eta\to 3\pi^{\circ}$ decay.
 (Right) radial density distribution (right) for the MC 
(shaded area) and experimental data (closed symbols).}
\end{figure}

This letter describes an exclusive measurement of kinematical
distributions of the $\pi^{\circ}$'s from the $\eta\to 3\pi^{\circ}$
channel. The phase space of the decay can be represented by a Dalitz
plot employing symmetrized variables as shown in
Fig.~\ref{fig:dalitzplot}(left). The amplitude can then be expressed
in terms of polar coordinates $\rho$ and $\phi$, where due to symmetry
$\phi$ is restricted to $0<\phi<60^\circ$.  Instead of $\rho$ a
normalized dimensionless variable $0\le z\le 1$ is used:
\begin{equation}
z\equiv \frac{\rho^2}{\rho^2_{max}}=
\frac{2}{3}\sum_{i=1}^3\left(\frac{T_i-<T>}{<T>}\right)^2.
\label{eqn:z}
\end{equation}
where $T_i$ is kinetic energy of a $\pi^{\circ}$ in the $\eta$ rest frame
and  $<T>$ is the average kinetic energy.

The slope parameter, $\alpha$,  is defined as a coefficient in 
the leading term of the decay amplitude expansion in $z$:
\begin{equation}
\mid A(z,\phi)\mid^2 =c_0( 1 + 2\alpha z).
\label{eqn:alpha}
\end{equation}  

The  situation for  the  $\alpha$ measurement  is  unsettled with  
inconsistent experimental results and disagreement with CHPT predictions
(Table~\ref{tab:table2}).  In lowest order CHPT $\alpha=0$. The
deviation is caused by the $\pi^{\circ}\pi^{\circ}$ interactions.  All
calculations beyond  one loop predict a negative value  of $\alpha$.
Dispersion  relations   \cite{Kambor:1995yc}  and   U(3)  chiral
effective     field     theory    with     rescattering
\cite{Beisert:2003zs} give consistently a low value for the $|\alpha|$
of  about  0.007.   Most   recent  calculations  with  U(3)  effective
Lagrangian   and  Bethe-Salpeter   equations   include  more   general
rescattering graphs  and parameters are fitted to the  
scattering data for the pseudoscalar mesons 
\cite{Borasoy:2005du}.    The  calculations   are   able  to   explain
the large experimental values for $|\alpha|$.  However in that approach
a change of the  $Q$ value  can be compensated  by other fit  parameters and
could even be equal to $Q_D$.

A special feature  of the $\eta\to 3\pi^\circ$  decay is that the
physical   region   extends  below   threshold   of  the   $\pi^0\pi^0
\to\pi^+\pi^-$ reaction.  Therefore a cusp structure should be visible
in invariant  mass distribution  of two $\pi^0$  around $2m_{\pi^\pm}$
(dashed lines in Fig.~\ref{fig:dalitzplot}(left)), in analogy to the
recent    observation     in    $K^+\to\pi^+\pi^0\pi^0$    decay
\cite{Batley:2005ax,Cabibbo:2004gq,    Belina:2006bb}.     The   exact
parameters of  the cusp  in the $K^+\to\pi^+\pi^0\pi^0$  decay provide
one of  most precise  determinations of the $\pi\pi$ scattering
length ($a_2-a_0$).

\begin{table}
\caption{\label{tab:table2} Experimental and theoretical results for the slope parameter $\alpha$}
\begin{ruledtabular}
\begin{tabular}{l|l|rl}
 $\alpha$ & Comment& Ref. &\\
\hline
-0.052 $\pm$ 0.017(stat) $\pm$ 0.010(syst)& Exp (CBarrel)  & \cite{Abele:1998yi}&                  \\
-0.031 $\pm$ 0.004                        & Exp (CBall)  & \cite{Tippens:2001fm}&                \\ 
-0.013 $\pm$ 0.004(stat) $\pm$ 0.005(syst)& Exp (KLOE)  & \cite{Giovannella:2005rz}&            \\
\hline
0                                          & PCAC  & \cite{PhysRevLett.18.1170}&                \\
+0.015                     & CHPT,1loop  & \cite{Gasser:1984pr,Bijnens:2002qy}&\\
-0.007...-0.014                            & CHPT$+$dispersive  & \cite{Kambor:1995yc}&                \\
-0.007                                    & UCHPT  & \cite{Beisert:2003zs}&                \\
-0.031 $\pm$ 0.003                        & UCHPT/fit  & \cite{Borasoy:2005ws}&            \\
\end{tabular}
\end{ruledtabular}
\end{table}

The   experimental  determination  of   the  $\alpha$   requires  that
background  contribution to  the $\eta$  production process  and other
systematic errors  are well under control.  All  recent experiments on
the  decay  $\eta \rightarrow  3\pi^{\circ}$  rely  solely on  photon
detection  with  close  to  4$\pi$ electromagnetic  calorimeters.   The
Crystal  Barrel  experiment  used the  final  state  with  10$\gamma$
($\overline{p}p  \to   \eta  2\pi^{\circ}$)  \cite{Abele:1998yi}, 
Crystal Ball used the reaction $\pi^-p  \rightarrow n\eta$ where only the 6$\gamma$
from   the    decay   were   measured    \cite{Tippens:2001fm}.    KLOE
\cite{Giovannella:2005rz} used the  decay $\phi\to\eta\gamma$ as  a source
of the $\eta$ mesons leading to  7$\gamma$ in the final state. 
The  $\alpha$  values  differ  by  more  than  three  standard
deviations  between the  experiments.  We  have  performed an independent
measurement using the reaction $pp\rightarrow pp\eta$ \cite{Pauly:2006pm}.
The main advantage  is that the excellent missing mass resolution  of the forward
going $pp$ pair can be used to tag the  process.

\section{Measurement and data reduction}

The analysis  is based  on data taken with  the WASA detector at
CELSIUS \cite{Zabierowski:2002ah} using  a pellet target system that
provides small  (30 $\mu$m diameter)  hydrogen pellets that  cross the
proton  beam of  nominal  1.36~GeV and  1.45~GeV  kinetic energy  (the
corresponding  center of mass  excess energies  are 41~MeV  and 75~MeV
respectively).   The  WASA detector  system  comprises a  multilayered
forward detector (FD) for  the measurement of charged particles emerging
in  the scattering  angle range  of 2.5$^{\circ}$-18$^{\circ}$,  and a
central deteector  (CD) composed of an  electromagnetic calorimeter of
1012 CsI(Na) crystals and a drift chamber/solenoid combination for the
measurement of  particles from $\eta$  decays in the angular  range of
20$^{\circ}-$140$^{\circ}$.

\begin{figure}
\includegraphics[width=0.5\textwidth]{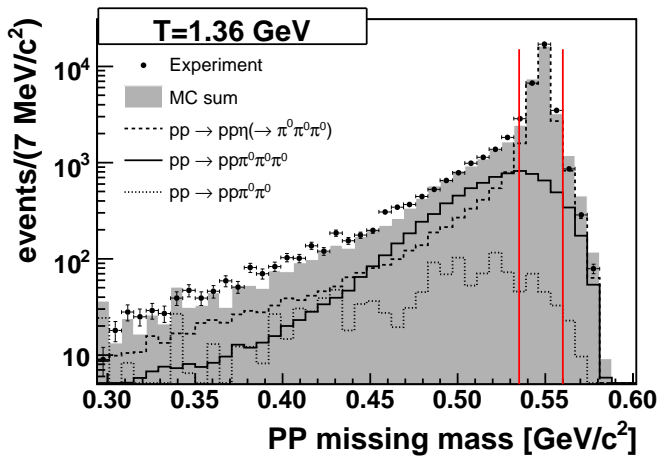}
\caption{\label{fig:PL1a} Experimental $pp$ missing mass distribution 
for reconstructed $pp\pi^{\circ}\pi^{\circ}\pi^{\circ}$ final states  
with a fit of 
the MC distributions for 2$\pi^{\circ}$, 3$\pi^{\circ}$ and 
$\eta$ production (from \cite{Pauly:2006pm}). The vertical 
lines indicate the $\eta$ selection. 
}
\end{figure}

The  basic criteria  for  the  selection  of  the  final  $\eta  \rightarrow
3\pi^{\circ}$ state  are (i) two protons  detected in the  FD and (ii)
six   $\gamma$    hit   clusters   (from    $\pi^{\circ}   \rightarrow
\gamma\gamma$) detected in the CD as neutrals with a minimum of 20 MeV
deposited  energy.   The   experimental  proton-proton  missing  mass
distribution can  be compared to  Monte Carlo (MC) simulated  data for
the  reaction  channels  $pp\to  pp\pi^{\circ}\pi^{\circ}\pi^{\circ}$,
$pp\to pp\eta$,  and $pp\to pp\pi^{\circ}\pi^{\circ}$.   The procedure
is described in detail in \cite{Pauly:2006pm}, the result for 1.36 GeV
kinetic energy is shown in  Fig. \ref{fig:PL1a}. The good agreement in
shape that  is obtained also for other  kinematical proton variables
and  of the absolute cross  sections for the  $\eta$ channel with
existing data,  demonstrates that the detector and  its efficiency are
well under control.   The missing mass resolution  of the FD
(5  MeV/c$^2$ FWHM  at 1.36  GeV)  therefore allows  selection of the   $pp\to
pp\eta$  reaction  (the condition  535 MeV/c$^2$ $<MM_{pp}<$560  MeV/c$^2$  was
applied) such that the remaining background, mostly from
direct $pp\to pp3\pi$ reaction, is about 5\%.

All possible combinations (15) of the 6 reconstructed gammas to form 3
$\pi^{\circ} \rightarrow \gamma \gamma$ pairs are sorted by means of a
parameter   $\chi^2   =   \sum{_{i=1}^3}{(IM_i  -   m_{\pi^\circ}
)^2}/{\sigma_{i}^2}$,  where  $\mathrm{IM}_i$ is  the  invariant  mass of  the
$i$-th pair  of a combination, and $\sigma_{i}$  the resolution.  At
most  two combinations  with the  lowest $\chi^2$  are selected  for a
kinematical fit  of the full  event with 8  constraints: four-momentum
conservation,  the  $  \pi^{\circ}\to\gamma  \gamma$  constraints  and
the $\eta\to 3\pi^{\circ}$ constraint.   A cut on the $\chi^2$  of the 
most probable combination
is  applied  to further  suppress  background  and to  increase the
combinatorial purity. Three different data sets were individually analyzed, the
combined result is based on 75000 events in the Dalitz plot after all cuts.

\section{Dalitz plot and slope $\alpha$}

  The reconstructed experimental $z$ distribution is shown in
Fig.~\ref{fig:dalitzplot}(right) together with MC simulation including
$pp\to pp\eta$ and $pp\to pp\pi^{\circ}\pi^{\circ}\pi^{\circ}$
reactions.  The detector response function for the reconstructed $z$
was determined from MC: it has a gaussian distribution with a standard
deviation (RMS) 0.067. The resolution in the $\pi^\circ\pi^\circ$
invariant mass ($M_{\pi\pi}$) is approximately constant and equal to 6
MeV/c$^2$ (RMS).
\begin{figure}
     \includegraphics[width=0.255\textwidth]{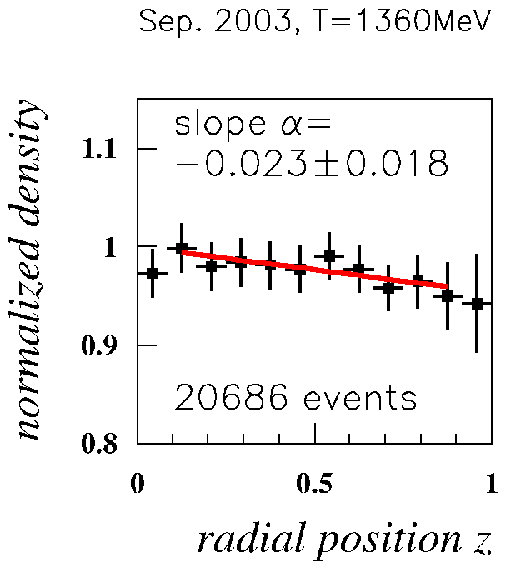}
     \includegraphics[width=0.217\textwidth]{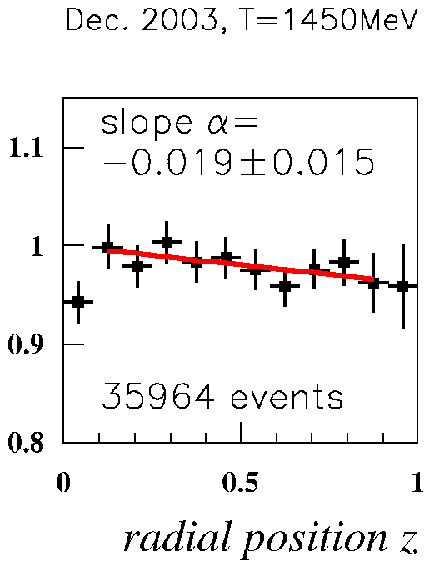}
 \caption{\label{fig:partialslopes}
Extracted experimental $|A|^2/c_0$ dependence on the $z$ variable 
for a subset of the T = 1.36 GeV (left) and for all 1.45 GeV (right) data  
together with a fit of the
slope parameter $\alpha$. Only statistical errors are shown.}
\end{figure}
\begin{table}
\caption{\label{tab:table1}Statistical errors of $\alpha$ fitted for data 
subsets obtained by 
parameter variation for selection 
criteria and reconstruction procedure.}
\begin{tabular}{l|c|c}
 Condition                                             & $\alpha \pm \sigma_{stat}$  & events \\ \hline
 All data, 0.1$\le z \le$ 0.9                          & -0.026 $\pm$ 0.010                   & 74700   \\
 All data, 0.0$\le z \le$ 0.9                          & -0.014 $\pm$ 0.009                   & 74700   \\
Analysis of subsets:                                                       &                                      &          \\
Subset I  T = 1.36 GeV                                 & -0.023 $\pm$ 0.018                   & 20700   \\
Subset II  T = 1.36 GeV                                & -0.041 $\pm$ 0.018                   & 18000   \\
Subset   T = 1.45 GeV                                  & -0.019 $\pm$ 0.015                   & 36000   \\
Subset I, variation of $MM_{pp}$ cut:                  &                                      &          \\
0.530 $\le MM_{pp}  \le$ 0.575                         & -0.019 $\pm$ 0.018                   & 21300    \\
0.535 $\le MM_{pp}  \le$ 0.560                         & -0.023 $\pm$ 0.018                   & 20700    \\
0.540 $\le MM_{pp}  \le$ 0.555                         & -0.025 $\pm$ 0.018                   & 18800    \\
Subset I, variation of $\chi^2_{kinfit}$ cut:          &                                      &          \\
$\chi^2_{kinfit} \le$ 999                              & -0.023 $\pm$ 0.015                   & 27900     \\
$\chi^2_{kinfit} \le$  50                              & -0.021 $\pm$ 0.017                   & 23600     \\
$\chi^2_{kinfit} \le$  30                              & -0.023 $\pm$ 0.018                   & 20700     \\
$\chi^2_{kinfit} \le$  15                              & -0.021 $\pm$ 0.020                   & 15200     \\
$\chi^2_{kinfit} \le$  15, $\chi^2_{second}$/$\chi^2_{first} \ge$ 1.2   & -0.038 $\pm$ 0.024  & 10800     \\
Subset I, alternative parameterization &&\\
of kin. fit     & -0.023 $\pm$ 0.019                   & 17000      \\
\end{tabular}
\end{table}
\begin{figure}
     \includegraphics[width=0.35\textwidth,clip]{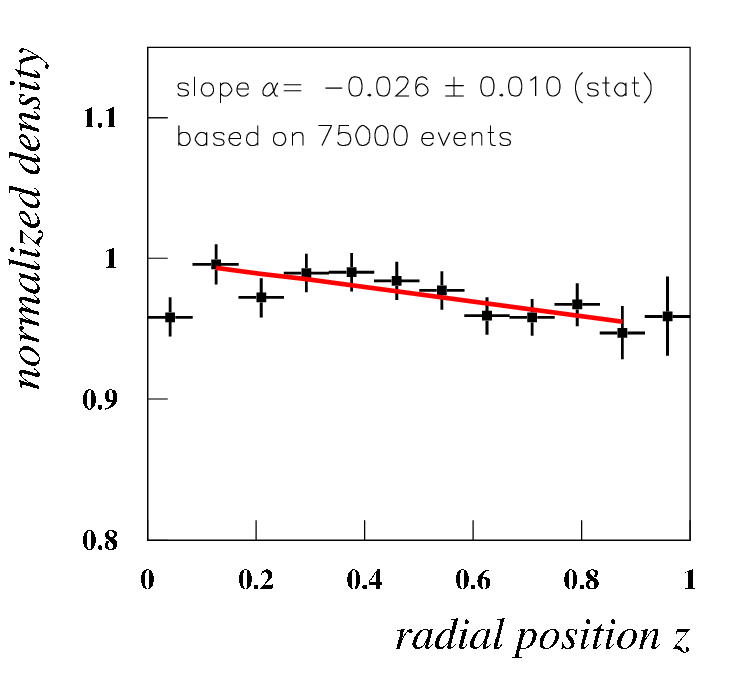}
     \caption{\label{fig:totalslope}
Extracted experimental $|A|^2/c_0$ dependence on the $z$ variable for all data together with a linear fit to 
of the slope parameter $\alpha$.  Only statistical errors are shown.}
\end{figure}

The  acceptance corrected $|A|^2$  dependence on  the $z$  variable is
obtained by  dividing the measured  distribution by the  MC prediction
with $\alpha$  set to 0.  A  fit of the  function (\ref{eqn:alpha}) is
performed  to  extract  the  parameter  $\alpha$.   The  MC  data  are
normalized to obtain  $c_0=1$.  For the final result  $\alpha = -0.026
\pm 0.01(stat) \pm 0.01(syst)$ the first bin was excluded due to large
systematic uncertainty  and the last  bin due to low  statistics.  The
estimate  of the  systematic errors  is obtained  by variation  of all
essential  cuts  applied  in  the  reconstruction:  $\chi^2$  for  the
kinematic  fit, the combinatorial  purity of  the sample,  the missing
mass and  the $z$  range.  For  each subsample a  fit to  $\alpha$ was
performed:  an  example  is shown  in Fig.~\ref{fig:partialslopes}  and  the
summary is given in Table~\ref{tab:table1}.  The overall systematical error was
obtained by  comparing central values  of the fits for  the subsamples
and  it takes  into account  the influence  of the  first bin  for the
$\alpha$ value~\cite{Pauly:2006zz}.

\begin{figure}
     \includegraphics[width=0.5\textwidth,clip]{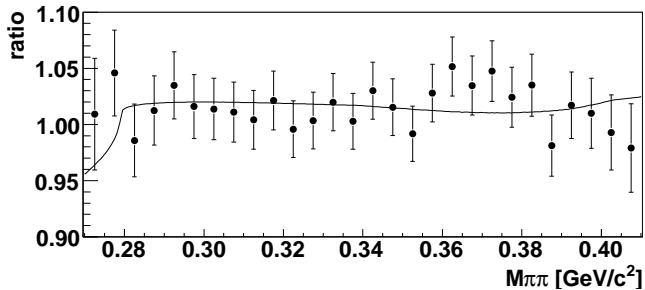}
     \caption{\label{fig:cusp}
Ratio of the $\mathrm{M}_{\pi\pi}$ distribution to
the MC predictions with $\alpha=0$: points -- data; line  -- CHPT calculations 
from reference \cite{Belina:2006bb}.}
\end{figure}

We have also performed a search for a cusp structure in the
invariant mass distribution of the two $\pi^\circ$
(Fig.~\ref{fig:cusp}). An eventual observation would require at least two
orders of magnitude larger data sample.

The  presented result for $\alpha$ is limited  by the  available  statistics which
determines also  attainable systematical accuracy (due to  the size of
the subsamples used for the tests). Within the errors it is compatible
both with the result from Crystal Ball \cite{Tippens:2001fm}, based on
10$^6$    events,    and   with    the    preliminary   KLOE    result
\cite{Giovannella:2005rz}   listed  in   Table~\ref{tab:table2}.   The
reconstruction methods  developed for  the purpose of  this experiment
are now  being used  in analysis of  recently collected data  from the
followup  experiment  with WASA  detector  located  at  COSY ring  (FZ
Juelich) \cite{Adam:2004ch} with much larger statistics.

\begin{acknowledgments}
We wish  to acknowledge the support  of the personnel  at The Svedberg
Laboratory throughout this experiment. This work has been supported by
BMBF  (06HH152, 06TU261),  by  Russian Foundation  for Basic  Research
(Grant  RFBR   02-02-16957),  and  the   European  Community  Research
Infrastructure Activity (FP6, Hadron Physics, RII-CT-2004-506078).
\end{acknowledgments}


\end{document}